\documentclass[aps,prb,twocolumn,floatfix,amsmath,amssymb,showpacs,
               superscriptaddress]{revtex4-1}
\usepackage{graphicx,placeins}
\usepackage{color}
\usepackage{dcolumn}
\usepackage{bm}
\usepackage{epsfig,psfrag,amsmath,amssymb,float}
\input{epsf}
\usepackage{hyperref}
\usepackage[percent]{overpic}
\hypersetup{
    colorlinks=true,
    linkcolor=blue,
    citecolor=blue,
    filecolor=magenta,      
    urlcolor=cyan,
}
\usepackage{array}
\newcolumntype{C}[1]{>{\centering\arraybackslash$}p{#1}<{$}}

\setcitestyle{square}
\newcommand{\pdag}{{\phantom{\dagger}}}
\bibpunct{[}{]}{,}{n}{}{}
\begin{document}
\title{BCS-BEC crossover in a $(t_{2g})^{4}$ Excitonic Magnet}

\author{Nitin Kaushal}
\affiliation{Department of Physics and Astronomy, The University of 
Tennessee, Knoxville, Tennessee 37996, USA}
\affiliation{Materials Science and Technology Division, Oak Ridge National 
Laboratory, Oak Ridge, Tennessee 37831, USA}

\author{Rahul Soni}
\affiliation{Department of Physics and Astronomy, The University of 
Tennessee, Knoxville, Tennessee 37996, USA}
\affiliation{Materials Science and Technology Division, Oak Ridge National 
Laboratory, Oak Ridge, Tennessee 37831, USA}

\author{Alberto Nocera}
\affiliation{Department of Physics and Astronomy and Stewart Blusson Quantum Matter Institute,
University of British Columbia, Vancouver, B.C., Canada, V6T 1Z1}

\author{Gonzalo Alvarez}
\affiliation{Computational Sciences and Engineering Division %
and Center for Nanophase Materials Sciences, Oak Ridge National Laboratory, %
Oak Ridge, Tennessee 37831, USA}

\author{Elbio Dagotto}
\affiliation{Department of Physics and Astronomy, The University of 
Tennessee, Knoxville, Tennessee 37996, USA}
\affiliation{Materials Science and Technology Division, Oak Ridge National 
Laboratory, Oak Ridge, Tennessee 37831, USA}

\date{\today}

\begin{abstract}
The condensation of spin-orbit-induced excitons in $t_{2g}^4$ electronic systems is attracting considerable attention. In the large Hubbard $U$ limit, 
antiferromagnetism was proposed to emerge from the Bose-Einstein Condensation (BEC) of triplons ($J_{\textrm{eff}}=1$). 
In this publication, we show that even for the weak and intermediate $U$ regimes, the spin-orbit exciton condensation is possible
leading also to staggered magnetic order. The canonical electron-hole excitations (excitons) transform into local triplon excitations at 
large $U$, and this BEC strong coupling regime is smoothly connected to the intermediate $U$ excitonic insulator region. We solved the degenerate 
three-orbital Hubbard model with spin-orbit coupling ($\lambda$) in one-dimensional geometry using the Density Matrix Renormalization Group, 
while in two-dimensional square clusters we use the Hartree-Fock approximation (HFA). Employing these techniques, we provide 
the full $\lambda$ vs $U$ phase diagrams for both one- and two-dimensional lattices. Our main result is that at the intermediate Hubbard $U$ region of our focus, 
increasing $\lambda$ at fixed $U$ the system transitions from an incommensurate spin-density-wave metal to a Bardeen-Cooper-Schrieffer (BCS) excitonic insulator,
with coherence length $r_{coh}$ of $\mathcal{O}(a)$ and $\mathcal{O}(10a)$ in $1d$ and $2d$, respectively, with $a$ the lattice spacing. 
Further increasing $\lambda$, the system eventually crosses over to the BEC limit (with $r_{coh} << a$). 
\end{abstract}
\maketitle

\section{Introduction}

Excitonic condensation has attracted considerable attention for over half a century, since its early theoretical prediction~\cite{rice1960}. An 
exciton is a bound state of an electron-hole pair. This composite particle resembles the Cooper pair of superconductors and follows hard-core bosonic 
statistics. Early work involving semiconductors showed that in the weak-coupling limit (small $U$), near the semimetal to semiconductor transition and 
depending on the excitonic binding energy and the band gap, the system can become unstable against the formation of multiple excitons~\cite{Mott01,Knox01} 
and this can lead to a condensation into a BCS-like macroscopic state called Excitonic Insulator. This BCS wavefunction smoothly transforms 
into a BEC state, if the gap between the bands increases at fixed Hubbard repulsion $U$. The phenomenon of excitonic condensation in 
strongly correlated systems can also lead to a BEC-like Excitonic Insulator in the strong coupling limit (large $U$). This regime 
also attracted considerable theoretical investigations ~\cite{Kunes01,Batista01,Sugimoto01} and has been studied using perturbation theory at 
large $U$, where the hopping amplitude $t$ is the small parameter. 

The extended Falicov-Kimball model has been used often as a minimal theoretical model to investigate 
the above discussed physics~\cite{Ejima01,Seki01,Phan01}. Only recently, the more realistic, but also more difficult, three-orbital Hubbard models were explored to study 
excitonic condensation~\cite{Sato02,exc2,Kaushal01,Kaushal02}. Due to the theoretical progress, on the experimental side there have been some studies showing reliable signatures of the excitonic condensate in real materials such as in a transition metal dichalcogenide~\cite{Kogar01}, Ta$_{2}$NiSe$_{5}$~\cite{YFLu01}, and in bilayer systems~\cite{Eisenstein01,Burg01}. To increase our understanding of this interesting Excitonic Insulator it is important to find additional candidate materials 
and additional theoretical models where the excitonic condensation occurs and can be studied in detail. Recently, it was recognized that materials with strong spin-orbit coupling can also provide a new avenue for the study of excitons and excitonic condensation~\cite{Cao02,Rau01}. For example, Sr$_{2}$IrO$_{4}$, with Ir$^{4+}$ ions and filling $n=5$ electrons per Iridium ion ($t_{2g}^{5}$), is a celebrated material due to the presence of long-range antiferromagnetic ordering in quasi two-dimensional layers, as in the case of superconducting cuprates~\cite{BJKim01}. Recent resonant inelastic X-ray scattering (RIXS) experiments on Sr$_{2}$IrO$_{4}$ have reported the presence of excitons as an excitation at approximately $0.5-0.6$ eV~\cite{Souri01,Kim02,Kim01}. RIXS experiments on one-dimensional stripes of Sr$_{2}$IrO$_{4}$ have also shown the presence of excitons at nearly the same energy~\cite{Gruenewald01}. These excitons are present in spin-orbit entangled states, hence known as \textit{spin-orbit excitons}.

Another promising avenue to study excitonic condensates induced by spin-orbit coupling involves the $t_{2g}^{4}$ case, 
which is the focus of the present paper. Theoretically, it has been predicted that the three-orbital Hubbard model 
with a degenerate $t_{2g}$ space and in the $LS$ coupling limit ($U\gg t,\lambda$) can lead to the BEC state of 
triplons (singlet-triplet excitations)~\cite{Khaliullin,meetei,Svoboda01}. We will show that these triplon excitations 
are low-energy manifestations of spin-orbit excitons. Cluster Dynamical Mean Field Theory (DMFT) calculations have also supported similar
findings~\cite{Sato02,exc2}, although it is difficult to distinguish between BCS and BEC states in the small clusters used. 
A recent computational study of the ground state of a one-dimensional spin-orbit coupled three-orbital Hubbard model in a non-degenerate (tetragonal) $t_{2g}$ space using the numerically-exact density matrix renormalization group (DMRG) also reported a phase with staggered spin-orbit excitonic correlations~\cite{Kaushal01}. All the above mentioned studies reveal an antiferromagnetic ordering accompanying the excitonic condensate. The $t_{2g}^{4}$ case is relevant for materials with Ir$^{5+}$ ions and other $4d/5d$ transition metal oxides with the same doping $n=4$. The presence of triplon condensation was initially discussed for double perovskite materials, such as Sr$_2$YIrO$_6$ and Ba$_2$YIrO$_6$ with a $5d^{4}$ configuration~\cite{Cao01,Dey01,Corredor01,Terzic01,Fuchs01,Nag01}, but later on RIXS experiments showed that the bandwidth of triplon excitations is not sufficiently large in comparison to $\lambda$ to develop condensation~\cite{Kusch01}. It should be remarked that recent RIXS experiments on Ca$_2$RuO$_4$ suggests that this compound could be a candidate material for excitonic magnetism~\cite{Gretarsson01}, and $ab-initio$ calculations have reached the same conclusion~\cite{Feld01}. 

To investigate the spin-orbit excitonic condensation, this paper uses a simple degenerate three-orbital Hubbard model. Using numerically 
exact DMRG simulations on one-dimensional chains, we show that even in the intermediate Hubbard repulsion regime ($U/W \approx 1$) 
an excitonic condensation is induced accompanied by antiferromagnetic ordering. This regime is stabilized by increasing sufficiently $\lambda$ starting at the 
incommensurate spin-density wave metallic region of $\lambda=0$. This numerical result of a spin-orbit excitonic condensate at intermediate $U/W$ cannot be understood 
using large $U$ perturbation theories. Moreover, in two-dimensional ($2d$) lattices, by using the Hartree-Fock Approximation (HFA), we have also found 
a similar excitonic insulator phase in both the weak and intermediate $U/W$ regimes. As our main result, we show that there is a BCS-BEC crossover inside the excitonic condensate phase, both in the $1d$ and $2d$ lattices we studied. The BCS limit of the excitonic insulator occurs at intermediate $U/W$  
(and also for weak $U/W$ in $2d$), and by increasing $\lambda$ (at fixed $U/W$) the BEC state is reached. The previously known BEC state present at large $U/W$, 
due to the condensation of triplons, is here shown to be smoothly connected to the BCS excitonic insulator of intermediate and weak $U$. At strong coupling, in the 
non-magnetic phase at large $\lambda$, higher than needed for the excitonic magnetic phase, these electron-hole pair excitations transform into 
low-energy triplons ($\Delta J_{\textrm{eff}}=1$) and higher-energy quintuplons ($\Delta J_{\textrm{eff}}=2$) excitations and those  triplons condense on decreasing $\lambda$. We provide 
numerical evidence for this triplon condensation using DMRG by calculating the excitonic pair-pair susceptibility. 
We also provide the full $\lambda$ vs $U$ phase diagrams for $1d$ and $2d$ lattices using the many-body techniques discussed above.

The organization of this paper is as follows. In Sec.~II, the model used and the computational methodology are presented. We discuss the atomic limit of the Hamiltonian in Sec. III. In Sec. IV, the main results are shown. In particular, in Sec.~IVA, the results on $1d$ lattices using the DMRG technique are presented. 
In Sec.~IVB, the results on $2d$ lattices using HFA are displayed to further support our main proposal of a 
BCS-BEC crossover in the model studied. In Sec. V, we discuss the overall results and present our conclusions.

\section{Model and Method}
For the study presented in this publication, we used a degenerate three-orbital Hubbard model. The Hamiltonian has three primary terms: the tight-binding kinetic energy, 
the standard on-site multiorbital Hubbard interaction, and the on-site spin-orbit coupling (SOC): $H = H_{K} + H_{\mathrm{int}} + H_{SOC}$. The tight-binding term is written as
\begin{equation}
H_{K} = \sum_{{\langle i ,j \rangle},\sigma,\gamma,\gamma^{\prime}}t_{\gamma\gamma^{\prime}}
(c_{{i}\sigma\gamma}^{\dagger}c^\pdag_{j\sigma\gamma^{\prime}}+\mathrm{H.c.}).
\end{equation}
The hopping amplitudes $t_{\gamma \gamma'}$ connect only nearest-neighbor lattice sites (in both the $1d$ chain and $2d$ square lattices).
As discussed earlier, here we use degenerate orbitals, that is, $t_{\gamma \gamma '} = \delta_{\gamma\gamma '}t$. In some cases we fixed $t=0.5$ for simplicity
but most of our results are expressed in terms of dimensionless ratios, such as $U/W$ or $\lambda/t$. 
The total bandwidth is $W=4.0|t|$ and $8.0|t|$ for the $1d$ and $2d$ lattices, respectively. The above mentioned orbitals -- labeled as 0, 1, and 2 -- 
could be associated respectively to the $d_{yz}$, $d_{xz}$, and  $d_{xy}$ orbitals, namely the $t_{2g}$ sector.
The on-site multiorbital Hubbard interaction term in the Hamiltonian consists of the standard several components
\begin{multline}\label{INT_term}
H_{\mathrm{int}} = U\sum_{{i},\gamma} n_{{i}\uparrow\gamma}
n_{{i}\downarrow\gamma} 
+\left(U'-J_{H}/2\right)\sum_{{i},\gamma<\gamma'} n_{{i}\gamma}
n_{{i}\gamma'} 
\\
  -2J_{H}\sum_{{i},\gamma<\gamma'} \mathbf{S}_{{i}\gamma} \cdot 
  \mathbf{S}_{{i}\gamma'} 
+J_{H}\sum_{{i},\gamma<\gamma'} \left( P^{\dagger}_{{i}\gamma} 
P_{{i}\gamma'} + \mathrm{h.c.} \right) . 
\end{multline}
In the equation above, $\mathbf{S}_{{i}\gamma}= {{1}\over{2}}\sum_{\alpha,\beta} 
c_{{i}\alpha\gamma}^{\dagger} \sigma_{\alpha\beta} c^\pdag_{{i}\beta\gamma}$ represents the spin at orbital $\gamma$ and lattice site ${i}$, 
while $n_{{i}\gamma}$ is the electronic 
density operator also at orbital $\gamma$  and site $i$. The first two terms are the intra- and inter-orbital electronic repulsion, respectively. 
The Hund coupling is contained in the third term, which favors the ferromagnetic alignment of the spins at different orbitals and the same site. Finally, 
the  $P_{{i}\gamma}=c_{{i}\downarrow\gamma}c_{{i}\uparrow\gamma}$ operator in the fourth term (pair-hopping) is the 
pair anhilation operator that arises from the matrix elements of the ``1/r'' Coulomb repulsion as in the early studies of Kanamori. 
We used the standard relation $U^{\prime}=U-2J_H$ due to rotational invariance, and we fix $J_{H}=U/4$ for all the calculations as employed widely in
previous efforts~\cite{Kaushal01,Kaushal02,Herbrych01,Patel01,Herbrych02}.

The SOC term is 
\begin{equation}\label{SO_term}
H_{\mathrm{SOC}}=\lambda\sum_{{i},\gamma,\gamma^{'},\sigma,\sigma^{'}}
{{\langle \gamma|{\bold{L}_{i}}|\gamma^{'}\rangle}\cdot{\langle\sigma|{\bold{S}_{i}}|\sigma^{'}\rangle}}
c_{i\sigma\gamma}^{\dagger}c_{i\sigma^{'}\gamma^{'}} \hspace{0.1cm},
\end{equation} 
where the coupling $\lambda$ is the SOC strength. 

Using the model described above, we performed calculations on one-dimensional chains employing 
the DMRG technique~\cite{White01,White02} for various system lengths, such as $L$ = 16, 24, and 32 sites. 
We have used up to 1600 states for the DMRG process and have maintained a truncation error below $10^{-8}$ throughout the finite 
algorithm sweeps. We used the corrected single-site DMRG algorithm~\cite{White03} with correction $a$ = 0.001-0.008, and performed 35 to 40 
finite sweeps to gain proper convergence depending on the system size. After this convergence,  we calculated the 
spin-structure factor $S(q)$, orbital-structure factor $L(q)$, excitonic momentum distribution function $\Delta_{m=1/2}(q)$, and 
coherence length $r_{coh}$. With the information gathered from all these observables, we constructed the phase diagram. To calculate spectral functions 
with the DMRG, we have used the correction vector method~\cite{Kuhner01}, with the Krylov-space approach~\cite{Nocera01}. We obtain the single-particle spectral 
function $A(q,\omega)$ and the excitonic pair-pair susceptibility $\Delta_{m=1/2}(q,\omega)$. These frequency-dependent observables require
heavy computational time, and multiple compute nodes. In our DMRG calculations, we target the total $J_{\textrm{eff}}^{z}$ of the system to reduce the 
computational cost~\cite{Kaushal01}. Details of the Hartree-Fock calculations are discussed in Sec.~IVB below.

\section{Atomic limit}

\begin{figure}[!h]
\hspace*{-0.52cm}
\vspace*{0cm}
\begin{overpic}[width=1.0\columnwidth]{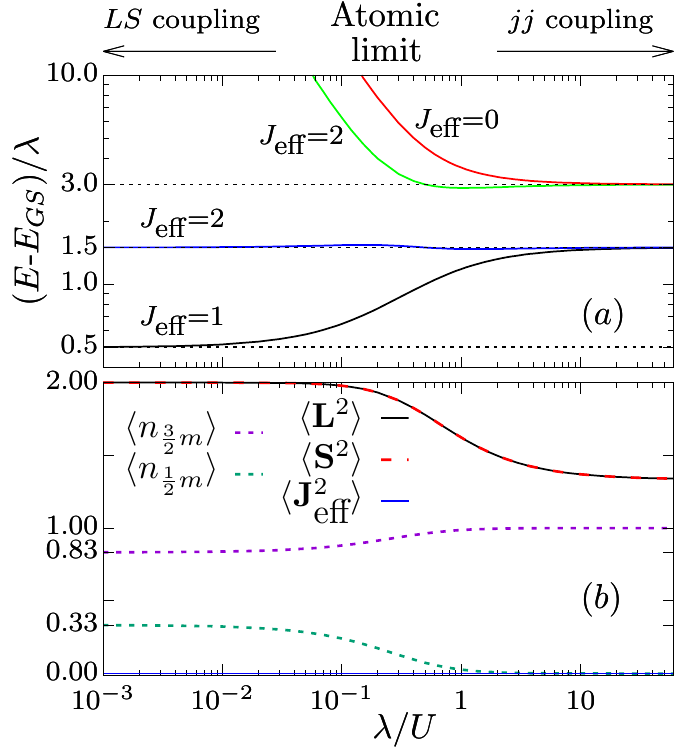}
\end{overpic}
\caption{Panel (a) shows the energies of excited states with respect to the ground state energy denoted by $E_{GS}$. In panel (b), 
the occupation in the ($j,m$) states and the local moments are shown. For the plots in this figure, $U=1$ is fixed and $\lambda$ varies.
The results illustrate the evolution from the $LS$ coupling to the $jj$ coupling regimes.}
\label{fig1}
\end{figure}

Before describing our results for chains and planes, we will discuss in detail the atomic limit of our Hamiltonian to introduce to the readers particular aspects of the 
$t_{2g}^4$ system. The magnetic properties in the $n=4$ case with a finite spin-orbit coupling are fascinating because in the atomic limit the 
ground state is a singlet having $J_{\textrm{eff}}=0$ for any finite value of $\lambda/U$. Only at $\lambda=0$, the $J_{\textrm{eff}}=0,1,2$ states 
are degenerate in the ground state manifold. Figure~\ref{fig1}(a) shows the energies of the excited states relative to the $J_{\textrm{eff}}=0$ ground state. The evolution of magnetic moments and occupations in the single-particle spin-orbit ($j_{\textrm{eff}},m$) states is displayed in Fig.~\ref{fig1}(b). Note that $j_{\textrm{eff}}$ is the effective total angular momentum of the electron and $m$ is the projection along the $z$-axis. Sometimes the quantum number $j_{\textrm{eff}}$ will be denoted as $j$, if it appears in subindex.


\begin{figure*}[!t]
\hspace*{-0.5cm}
\vspace*{0cm}
\begin{overpic}[width=2.1\columnwidth]{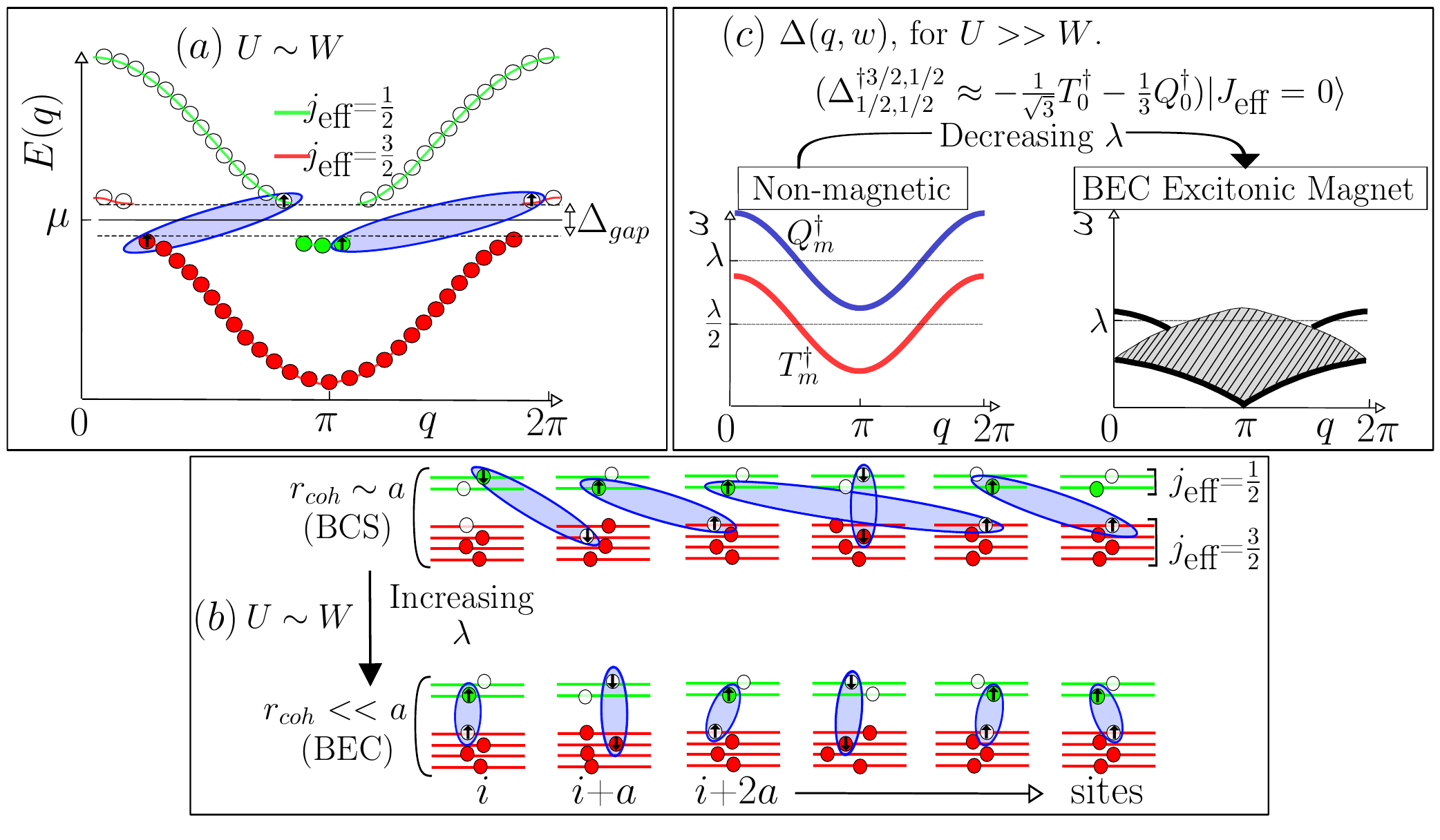}
\end{overpic}
	\caption{Visual representation of the main results of this publication, all supported by actual DMRG and Hartree Fock calculations. 
In (a), the single-particle excitations of the $j_{\textrm{eff}}=3/2$ and $j_{\textrm{eff}}=1/2$ bands are shown at intermediate $U$, where near the chemical potential 
the gap opens due to the formation of excitons (note electron and hole have the same $m$). The $j_{\textrm{eff}}=3/2$ and $j_{\textrm{eff}}=1/2$ bands open gaps 
near momentum $q\approx 0$ and $q\approx \pi$, respectively. In (b), the real-space perspective of the 
excitonic state (at intermediate $U$) is shown, where the exciton's mean radius (characterized by the coherence length) decreases by increasing $\lambda$. 
In (c), the excitonic condensation mechanism in the strong coupling limit is depicted. The local exciton creation operator leads to the creation of both a 
triplon and a quintuplon when acting on $|J_{\textrm{eff}}=0 \rangle$. In the presence of kinetic energy, i.e. including the tight-binding term,  
the triplon and quintuplon excitations gain bandwidths, and decreasing $\lambda$ leads to the Bose-Einstein condensation of the triplon component.}
\label{fig2}
\end{figure*}

The $jj$ coupling limit ($\lambda/U>>1$) can be understood by diagonalizing the spin-orbit coupling term.
The transformation between the $t_{2g}$ orbitals and the $j_{\textrm{eff}}$ basis is given by (site $i$ index dropped)
\begin{equation}\label{transformation}
\renewcommand{\arraystretch}{1.5}
\begin{bmatrix}a_{\frac{3}{2},\frac{3s}{2}}\\a_{\frac{3}{2},-\frac{s}{2}}\\a_{\frac{1}{2},-\frac{s}{2}}\end{bmatrix}
= \begin{bmatrix}\frac{is}{\sqrt{2}}&\frac{1}{\sqrt{2}}&0\\\frac{s}{\sqrt{6}}&\frac{i}{\sqrt{6}}&\frac{2}{\sqrt{6}}\\
\frac{-s}{\sqrt{3}}&\frac{-i}{\sqrt{3}}&\frac{1}{\sqrt{3}}
\end{bmatrix}\begin{bmatrix}c_{\sigma yz}\\c_{\sigma xz}\\c_{\bar{\sigma} xy}\end{bmatrix} ,
\end{equation}
where $s$ is $1(-1)$ when $\sigma$ is $\uparrow(\downarrow)$ and $\bar{\sigma}=-\sigma$. The $H_{SOC}$ term in the $j_{\textrm{eff}}$ basis becomes
\begin{eqnarray}\label{H_SO}
H_{\mathrm{SOC}}&=&\sum_{{i}}\frac{\lambda}{2}(-a_{{i},\frac{3}{2},\frac{3}{2}}^{\dagger} a_{{i},\frac{3}{2},\frac{3}{2}}^{\phantom{\dagger}}
- a_{{i},\frac{3}{2},-\frac{1}{2}}^{\dagger} a_{{i},\frac{3}{2},-\frac{1}{2}}^{\phantom{\dagger}}
\nonumber\\
&-&a_{{i},\frac{3}{2},-\frac{3}{2}}^{\dagger} a_{{i},\frac{3}{2},-\frac{3}{2}}^{\phantom{\dagger}}
-a_{{i},\frac{3}{2},\frac{1}{2}}^{\dagger} a_{{i},\frac{3}{2},\frac{1}{2}}^{\phantom{\dagger}}
\nonumber\\
&+&2a_{{i},\frac{1}{2},\frac{1}{2}}^{\dagger} a_{{i},\frac{1}{2},\frac{1}{2}}^{\phantom{\dagger}} 
+ 2a_{{i},\frac{1}{2},-\frac{1}{2}}^{\dagger} a_{{i},\frac{1}{2},-\frac{1}{2}}^{\phantom{\dagger}})\,.
\end{eqnarray}
The expression above clearly shows that for $n=4$ the ground state has all $j_{\textrm{eff}}=3/2$ states fully occupied, 
and all $j_{\textrm{eff}}=1/2$ states empty (i.e. $|GS\rangle_{jj}=a_{\frac{3}{2},\frac{-3}{2}}^{\dagger}a_{\frac{3}{2},\frac{3}{2}}^{\dagger}a_{\frac{3}{2},\frac{-1}{2}}^{\dagger}a_{\frac{3}{2},\frac{1}{2}}^{\dagger}|0\rangle$) leading to a total $J_{\textrm{eff}}=0$. 
The first excited state is located with a gap $3\lambda/2$, as shown in Fig.~\ref{fig1}(a). Also it can be checked that       
 $_{jj}<GS|\bold{L}^{2}|GS>_{jj}=_{jj}<GS|\bold{S}^{2}|GS>_{jj}=4/3$. On the other hand, 
in the $LS$ coupling limit  only one-third fraction of the ($j_{\textrm{eff}}=1/2,m$) states is occupied, 
with both the magnitudes of S and L being 1.
 
The information presented above in the atomic limit is relevant for cases with $U/W \ll 1$ and a finite spin-orbit coupling, 
to find out in which limit the system is located. For example, recently compounds with isolated RuCl$_{6}$ octahedra 
exhibited a single-ion physics with a non-magnetic $J_{\textrm{eff}}=0$ state~\cite{Lu01}. Next, we will discuss our results 
for the $1d$ and $2d$ lattices where the presence of kinetic energy can drive the system away from this atomic limit, leading to magnetic ordering.

\section{DMRG results in one dimension}

\begin{figure*}[!t]
\hspace*{-0.52cm}
\vspace*{0cm}
\begin{overpic}[width=2.1\columnwidth , height=0.9\columnwidth]{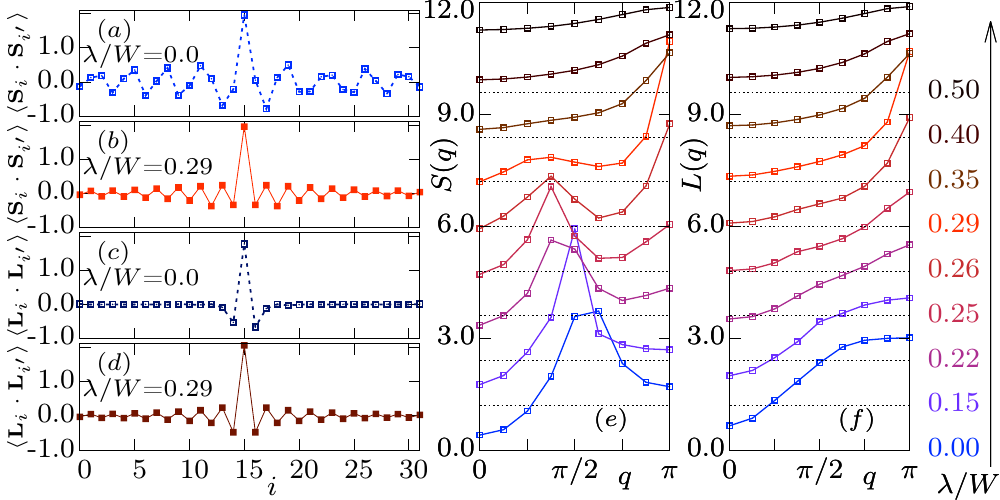}
\end{overpic}
\caption{DMRG results at fixed $U/W=2$. Panels (a) and (b) show the real-space spin-spin correlations at $\lambda/W=0$ and $\lambda/W=0.29$, repectively. 
In (c) and (d), the real-space orbital-orbital correlations are shown at $\lambda/W=0.0$ and $\lambda/W=0.29$, respectively. The spin structure factor $S(q)$ and orbital structure factors $L(q)$ are shown in panels (e) and (f), respectively, for various values of $\lambda$. For panels (a,b,c,d) the system size was $L=32$ 
and one site is fixed at the center i.e. $i^{'}=15$. A system size $L=16$ is used for panels (e,f).}
\label{fig3}
\end{figure*}

This section presents the DMRG numerical results for one-dimensional chains, and discuss their implications.
Figure~\ref{fig2} visually summarizes our conclusions. Within our numerical accuracy, 
we observed that the excitonic condensation starts at intermediate Hubbard correlation strength $U \approx\mathcal{O}(W)$. 
This condensation of excitons opens a gap in the $j_{\textrm{eff}}=3/2$ and $j_{\textrm{eff}}=1/2$ bands at momentum $q=0$ and $q=\pi$, 
respectively, as shown in Fig.~\ref{fig2}(a). A similar perspective for the gap opening near the Fermi level by excitonic condensation 
was discussed in early research for semiconductors~\cite{rice1960}. At intermediate $U$, we noticed the excitons condense at finite 
momentum $\pi$ and in the triplet channel (for details see Supplementary~\cite{splm}). 
We also noticed that increasing $\lambda$ (concomitantly adjusting $U$ to remain inside the
excitonic condensation region) decreases the coherence length ($r_{coh}$) of electron-hole pairs from approximately 
one lattice spacing ($a$) to a much smaller number $r_{coh}<<a$, resembling the BCS-BEC crossover. 
Although in the extreme BCS limit $r_{coh}$ can be as large as hundreds of lattice spacings, outside our resolution, 
in our finite and one dimensional system we only found a robust indication for a  maximum $r_{coh}$ of approximately $1.0a$ 
which definitely is different from the BEC limit $r_{coh}<<a$.  Confirming that indeed we found a BCS-BEC crossover at intermediate $U$, 
we performed mean-field calculations on $2d$ lattices (Sec.~V), where we found $r_{coh}$ as large as $\mathcal{O}(10a)$ in the BCS limit. 

We have also found clear distinctions between the exciton-exciton correlation 
decay in the BCS and BEC limits in one-dimensional chains.
It can be shown that the excitonic operator $\Delta_{1/2,m}^{\dagger3/2,m^{'}}$ in the single-atom $LS$ coupling limit can 
be written in terms of triplon and quintuplon excitations (see Supplementary~\cite{splm}), corresponding, respectively, to $\bold{T}_{n}^{\dagger}|J_{\textrm{eff}}^{z}=0, \bold{J_{\textrm{eff}}}=0\rangle=|n,1\rangle$, and $\bold{Q}_{l}^{\dagger}|J_{\textrm{eff}}^{z}=0, \bold{J_{\textrm{eff}}}=0\rangle=|l,2\rangle$, where $n \in \{\pm 1,0\}$ and $l \in \{\pm 2, \pm 1, 0 \}$. By calculating the pair-pair susceptibity of excitons $\Delta(q,\omega)$ for one-dimensional chains, in the non-magnetic phase present in the strong coupling limit, we found bands of triplon and quintuplon excitations, with minima at $q=\pi$ but both bands are gapped. We noticed 
that decreasing $\lambda$ drives the system to the BEC state, where $\Delta(q,\omega)$ has clear signatures of gapped triplon and quintuplon excitations 
only near $q=0$ but mainly a continuum of excitations at other momenta above the Goldstone-like modes emerging from $q=\pi$, as sketched in Fig.~\ref{fig2}(c).

\subsection{Magnetic properties and staggered excitonic correlations}

Now let us discuss the results for magnetism in one-dimensional chains. We choose $U/W=2.0$ and $U/W=10.0$ 
as representative points for the intermediate and strong coupling regions, respectively. First, consider the intermediate coupling region 
where at $\lambda=0$ we found an incommensurate spin-density wave (IC-SDW) via 
the spin-spin correlation $\langle \bold{S}_{i}\cdot\bold{S}_{{i}'}\rangle$ together with an exponential fast decay in the orbital-orbital 
correlation $\langle \bold{L}_{i}\cdot\bold{L}_{{i}'}\rangle$, as shown in Figs.~\ref{fig3}(a,c).  Block magnetic states, 
as discussed in~\cite{Herbrych01} and~\cite{Herbrych02}, do not appear in our model at $\lambda=0$.
Increasing $\lambda$ drives 
the system towards antiferromagnetism with staggered spin-spin and orbital-orbital correlations, as shown in Figs.~\ref{fig3}(b,d). 
For additional confirmation, we show the spin structure factor $S(q)=(1/L)\sum_{i,j}e^{\iota q(j-i)}\langle \bold{S}_{i}\cdot\bold{S}_{j} \rangle$, 
and the orbital structure factor $L(q)=(1/L)\sum_{i,j}e^{\iota q(j-i)}\langle \bold{L}_{i}\cdot\bold{L}_{j} \rangle$ for various $\lambda$ values. 
As $\lambda$ increases, the spin structure factor $S(q)$ in Fig.~\ref{fig3}(e) indicates that the incommensurate peak 
shifts to lower $q$ values. The antiferromagnetic tendencies, shown by the $q=\pi$ peak, starts only near $\lambda=0.22W$, and on further 
increasing $\lambda$ the $q=\pi$ peak grows and the incommensurate peak is reduced. At larger $\lambda$'s the $S(q=\pi)$ peak decreases as the system 
transitions into the non-magnetic state. The orbital structure factor $L(q)$ displays similar behaviour at $q=\pi$. $L(q)$ starts with a flat plateaux 
near $q=\pi$, then $L(q=\pi)$ grows when increasing $\lambda$, and eventually the $L(q=\pi)$ peak vanishes for very large $\lambda$, 
as shown in Fig.~\ref{fig3}(f).

\begin{figure}[!t]
\hspace*{-0.52cm}
\vspace*{0cm}
\begin{overpic}[width=1.0\columnwidth , height=1.2\columnwidth]{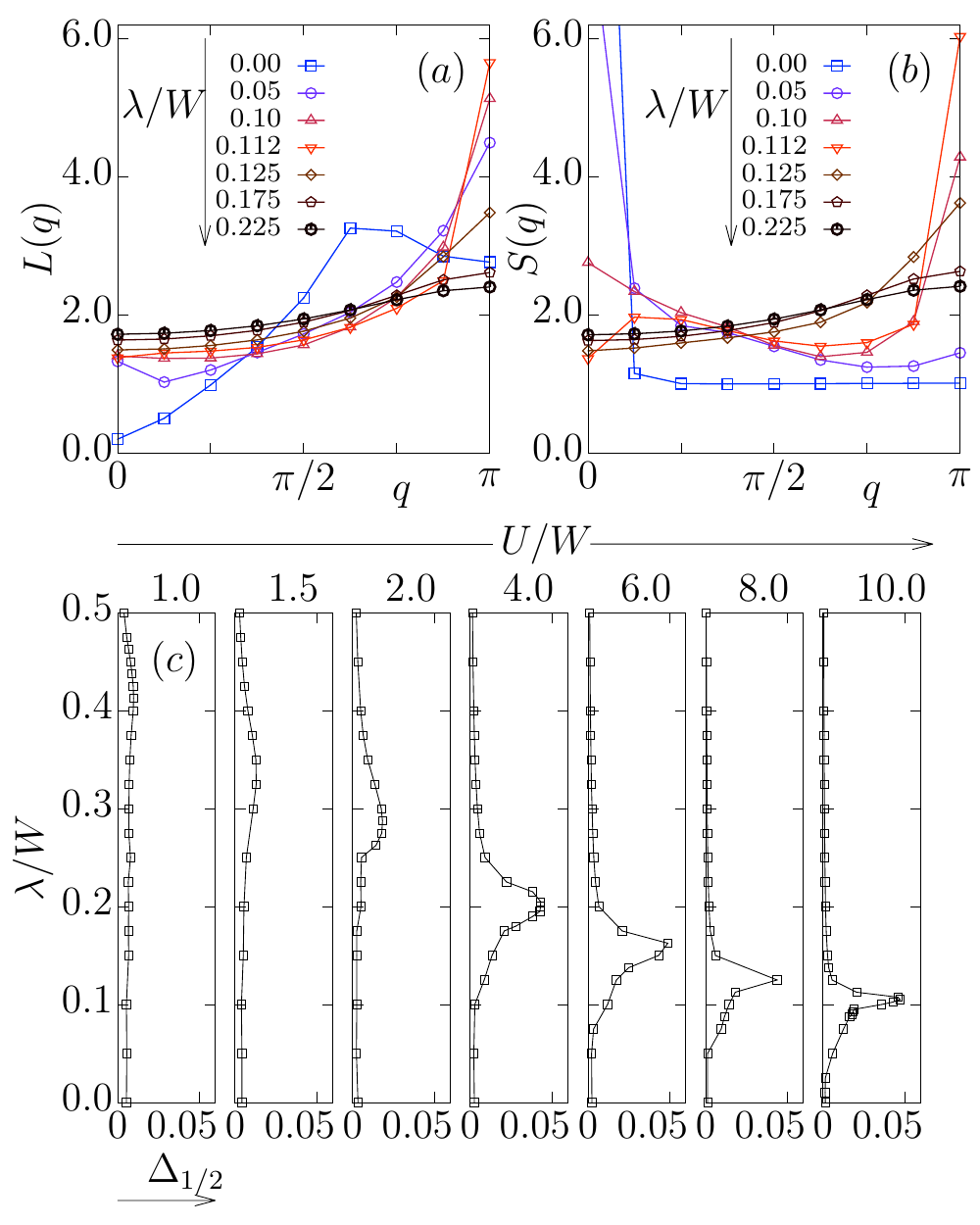}
\end{overpic}
\caption{Results calculated using DMRG for a one-dimensional chain of $L=16$ sites. In panels (a) and (b), the orbital structure factor $L(q)$
and spin structure factor $S(q)$ are shown, respectively, at fixed $U/W=10$ and for various $\lambda/W$'s. 
Panel (c) shows the measure of staggering in excitonic correlations $\Delta_{1/2}$ for various values of $\lambda$ and $U$.}
\label{fig4}
\end{figure}

In the strong coupling limit, and at $\lambda=0$, the spins align ferromagnetically and orbital-orbital correlations show a ``$+- -+- -$" pattern, 
depicted by peaks at momentum $q=0$ and near $q=2\pi/3$ in $S(q)$ and $L(q)$, respectively (see Fig.~\ref{fig4}). Note that recent DMRG 
calculations on the low-energy S=1 and L=1 model also showed similar results in the orbital correlations~\cite{Feng01} at $\lambda=0$. 
We noticed that as $\lambda$ increases, the system enters into the phase where orbital ordering becomes staggered, as shown by 
a peak at $q=\pi$ for $\lambda=0.05W$ in Fig.~\ref{fig4}(a). However, the spin ordering is dominantly ferromagnetic with small antiferromagnetic 
tendencies leading to a small peak at $q=\pi$ [Fig.~\ref{fig4}(b) for $\lambda=0.05W$]. Further increasing $\lambda$, both $L(q=\pi)$ 
and $S(q=\pi)$ grow while $S(q=0)$ decreases, and ultimately $L(q=\pi)$ and $S(q=\pi)$ also start decreasing when the system enters 
into the non-magnetic phase with exponentially decaying spin and orbital correlations. 

In our DMRG calculations, we noticed that the development of antiferromagnetic correlations in the spin and orbital channels 
is always accompanied by staggering in the exciton-exciton correlations,
both in the intermediate and strong coupling regions. We estimate the amount of staggering in excitonic correlation for our 
one-dimensional chains using:
\begin{equation}
\Delta_{m} = \frac{1}{L^{2}} \sum_{|i-i'|>0}(-1)^{|i-i'|}\langle \Delta_{1/2,m}^{\dagger 3/2,m}(i) \Delta_{1/2,m}^{3/2,m}(i') \rangle ,
\end{equation}
 where $m \in \{ \pm1/2 \}$. In Fig.~\ref{fig4}(c), the evolution of $\Delta_{1/2}$ is shown when changing $\lambda$, where
each panel belongs to a different $U/W$. 
 We found that smaller $\lambda$'s are needed as $U/W$ increases to obtain staggered excitonic correlations. 
A finite range of $\lambda$ where the excitonic correlations are staggered is present for $U/W$ as low as 1.5, although we noticed 
that the magnitude of $\Delta_{1/2}$ decreases as $U/W$ decreases, and for $U/W \lessapprox 1.0$ we were unable to identify -- within our numerical accuracy --  the region 
with staggered excitonic correlations. Nonetheless, it is interesting to note that we find  staggered excitonic correlations at intermediate 
$U$, where small $\lambda$ shows IC-SDW metal (for evidence of metallicity see Sec.~V.C).
Perturbation theories developed in the limit $U\gg t,\lambda$ cannot explain these results.

\subsection{BCS-BEC crossover, and IC-SDW metal to BCS-Excitonic insulator transition}
\begin{figure}[!t]
\hspace*{-0.52cm}
\vspace*{0cm}
\begin{overpic}[width=1.0\columnwidth , height=1.3\columnwidth]{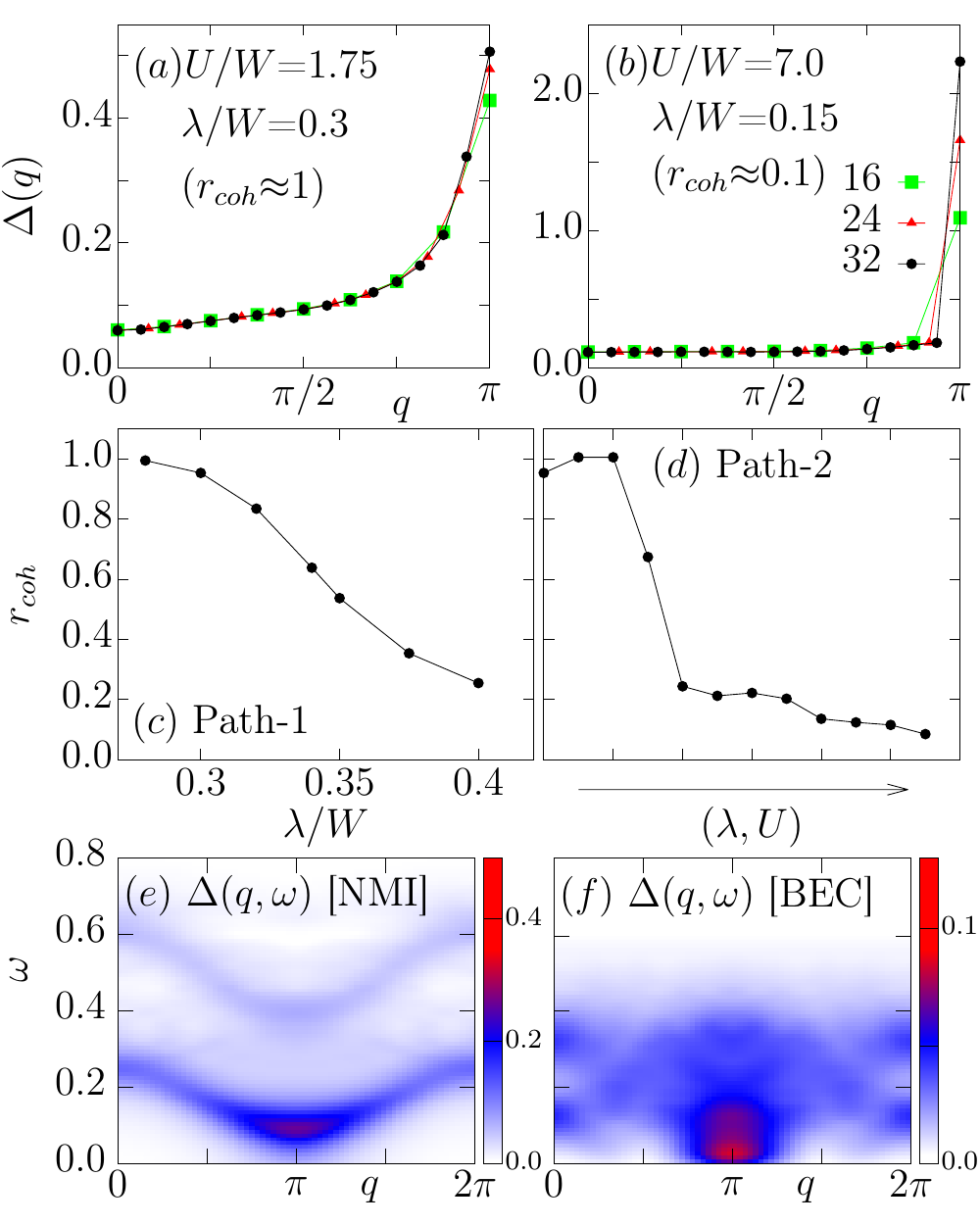}
\end{overpic}
\caption{Panels (a) and (b) show the momentum distribution function of excitons $\Delta(q)$ for system sizes $L=16,24,32$ 
at two representative points of the BCS and BEC regions, respectively. Panels (c) and (d) show the coherence length for two paths 
inside the excitonic condensate regime, shown in the phase diagram Fig.~\ref{fig7}. The pair-pair susceptibilities $\Delta(q,\omega)$ 
are shown in panel (e) at $U/W=10$ and  $\lambda/W=0.2$ i.e. in the non-magnetic insulator (NMI) region, and in panel (f) 
at $U/W=10$ and $\lambda/W=0.11$ i.e. in the BEC region. All the calculations above were obtained using DMRG.}
\label{fig5}
\end{figure}
\begin{figure*}[!t]
\hspace*{-0.5cm}
\vspace*{0cm}
\begin{overpic}[width=2.15\columnwidth]{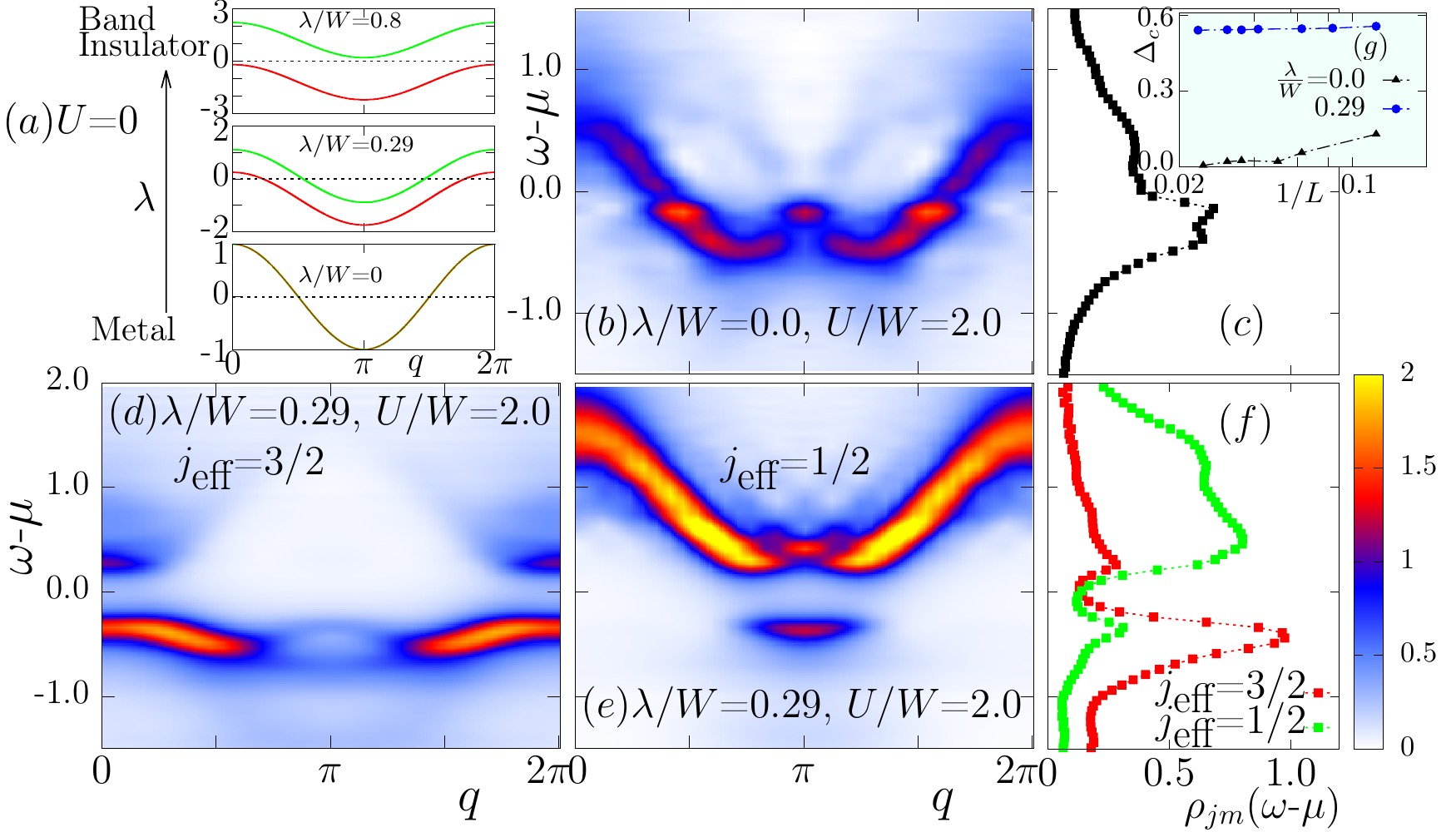}
\end{overpic}
\caption{The single-particle spectral function calculated using DMRG. The non-interacting band structure is shown in panel (a). 
Panel (b) contains $A_{jm}(q,\omega)$ for $\lambda=0$ and $U/W=2.0$, in the IC-SDW metallic phase. In panels (d,e) $A_{jm}(q,\omega)$ 
at $\lambda/W=0.29$ and $U/W=2.0$ is shown inside the excitonic insulator phase ($r_{coh}=1.019$). Panels (c) and (f) show the 
single-particle DOS ($\rho_{jm}(\omega)$) at $\lambda=0.0$ and $\lambda=0.29$, respectively. The scaling of the charge gap is shown 
in inset (g) for ($U/W=2, \lambda/W=0$) and ($U/W=2, \lambda/W=0.29$). }
\label{fig6}
\end{figure*}
\begin{figure}[!h]
\hspace*{-0.52cm}
\vspace*{0cm}
\begin{overpic}[width=1.0\columnwidth , height=0.6\columnwidth]{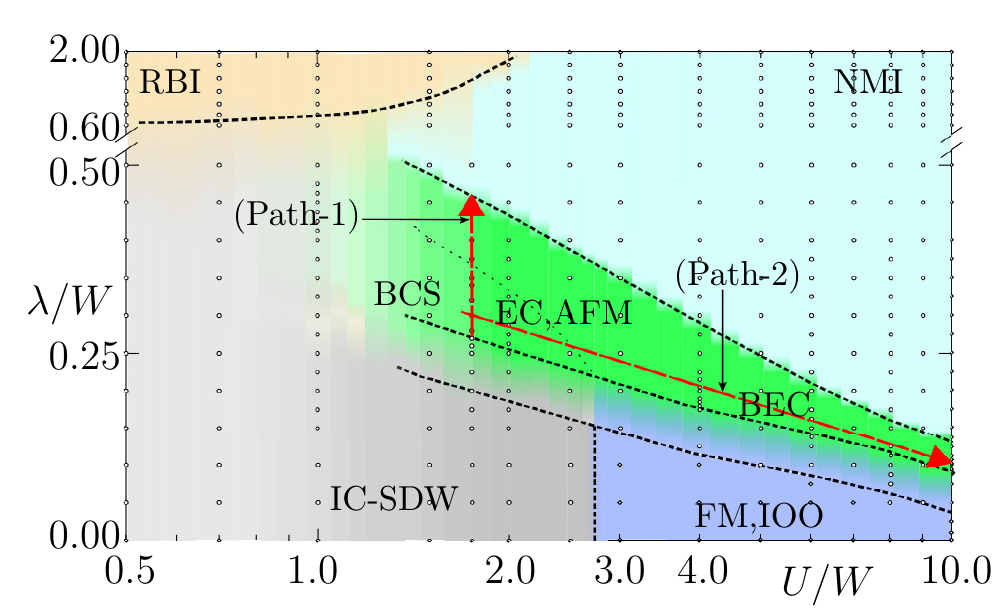}
\end{overpic}
\caption{$\lambda/W$ vs $U/W$ phase diagram calculated using DMRG for a one-dimensional system. The two red arrows corresponds to the paths used in panels (c,d) of Fig.~\ref{fig5}. The vertical red arrow is choosen at $U/W=1.75$, and depicts Path-1 of Fig.~\ref{fig5}(c). The diagonal red arrow corresponds to the Path-2 used in Fig.~\ref{fig5}(d). The notation RBI, IC-SDW, FM, IOO, EC, AFM, and NMI stands for relativistic band insulator, incommensurate spin-density wave, ferromagnetic, excitonic condensate, antiferromagnetic, and non-magnetic insulator, respectively.}
\label{fig7}
\end{figure}

Now we will discuss the main result of this paper, where we present the numerical evidence for the BCS-BEC crossover in the 
excitonic condensate region. At intermediate $U/W$ -- the value $U/W=1.75$ is chosen for this discussion merely for simplicity -- 
there is a finite range of $\lambda/W \in \{0.28, 0.4 \}$  where staggering in excitonic, spin, and orbital correlations 
is present. We calculate the coherence length $r_{coh}(m)$ using the widely-employed formula~\cite{Kunes01,Kaneko01,Kaneko02}
\begin{equation}
r_{coh}(m)=\sqrt{\frac{\sum_{ij} |i-j|^2\langle a_{i\frac{1}{2}\frac{m}{2}^{\dagger}}a_{j\frac{1}{2}\frac{m}{2}} \rangle } {\sum_{ij} \langle a_{i\frac{1}{2}\frac{m}{2}^{\dagger}}a_{j\frac{1}{2}\frac{m}{2}} \rangle }},
\end{equation}
where $m \in \{ \pm1/2 \}$, for the points lying inside the excitonic condensate region at fixed $U/W=1.75$, namely the Path-1 shown in 
Fig.~\ref{fig7}. Note that $r_{coh}(1/2)=r_{coh}(-1/2)$. Interestingly, we found that as $\lambda$ increases, $r_{coh}$ decreases 
from nearly one lattice spacing $\approx a$ to 
$\approx 0.2a$, see Fig.~\ref{fig5}(c). This reduction in the coherence length of electron-hole pairs resembles the BCS-BEC crossover 
already discussed in the context of semiconductors near the semi-metal to semiconductor transition~\cite{rice1960,Mott01,Knox01}. We also 
calculate $r_{coh}$ for various values of $\lambda$ and $U$ when transitioning from the intermediate to the strong coupling limits 
while being still inside the excitonic condensate region, which we call Path-2 (see Fig.~\ref{fig7}), as shown in Fig.~\ref{fig5}(d). Here 
again we found that $r_{coh}$ decreases from $\mathcal{O}(a)$ to $\mathcal{O}(0.1a)$. The ordered $(\lambda/W,U/W)$ points choosen for Path-2 
are $\{(0.3,1.75)$, $(0.29,2.0)$, $(0.29, 2.25)$, $(0.28,2.5)$, $(0.24, 3.0)$, $(0.23, 3.5)$, $(0.21, 4.0)$, $(0.2,4.5)$, $(0.17, 6.0)$, $(0.15,7)$, $(0.13,9)$, $(0.11,10)\}$. 

In Fig.~\ref{fig5}(a,b) we show the excitonic momentum distribution function $\Delta(q)=(1/L)\sum_{i,j}\langle e^{\iota q(i-j)} \Delta_{1/2,m}^{\dagger 3/2,m}(i) \Delta_{1/2,m}^{3/2,m}(j) \rangle$ for 
two points ($\lambda/W=0.3$, $U/W=1.75$) and ($\lambda/W=0.15$, $U/W=7.0$) corresponding to the intermediate $U$ BCS and strong $U$ BEC regions. 
We found that in the BCS limit $\Delta(q=\pi)$ grows very slightly with system size $L$, which implies the real-space correlations of local excitons have exponential decay in this regime (see Supplementary~\cite{splm}). On the other hand, in the BEC region we found nearly linearly increasing $\Delta(q=\pi)$ with $L$, which implies 
either a very slow power-law decay or even true long-range order. Such a true long-range order in our one-dimensional system is allowed as 
the $U(1)$ symmetry related to the excitonic condensation is explicitly broken by a finite Hund coupling~\cite{Kaushal02}. The analysis 
above also clearly implies that as we increase the system size and hence increase the number of excitons, these excitons can condense also
at momentum $q\ne \pi$ in the BCS limit, whereas in the BEC limit excitons condense only at $q=\pi$. For completeness, we would like 
to mention that a similar BCS-BEC crossover has also been reported in the extended Falicov-Kimball model in one-dimensional chains~\cite{Ejima01}.

As discussed before, the exciton creation (electron-hole pair excitation) becomes the triplon and quintuplon excitation in the atomic 
$LS$ coupling limit. We calculated the excitonic pair-pair susceptibility 
\begin{equation}
\Delta(q,\omega)=\langle\Psi_{0}|  \Delta_{1/2,1/2}^{\dagger3/2,1/2}(q) \frac{1}{\omega+i\eta - H + E_{0}} \Delta_{1/2,1/2}^{3/2,1/2}(q) |\Psi_{0}\rangle
\end{equation}
 in the strong coupling limit $U/W=10.0$ on one-dimensional chains to study the effect 
of the kinetic energy. The broadning $\eta$ was fixed at 0.05eV. We choose two values $\lambda/W=0.20$ and $\lambda/W=0.11$ in the non-magnetic insulator and BEC regions, 
respectively [Figs.~\ref{fig5}(e,f)]. We found that for the non-magnetic state, the pair-pair susceptibility shows two cosine-like bands 
with minima at $q=\pi$, where the lower band belongs to $\Delta J_{\textrm{eff}}=1$ (triplon) and the upper band represent 
$\Delta J_{\textrm{eff}}=2$ excitations, in agreement with previous analytical studies~\cite{Svoboda01}. The BEC is expected 
to occur by reducing $\lambda$, when the lower band of triplons becomes gapless. In our BEC state, $\Delta(q,\omega)$ shows features very different  
from those in the non-magnetic state: after removing the elastic peak we found that two gapped bands appear only near $q=0$, and the 
spectrum at $q=\pi$ is now gapless with emerging Goldstone-like modes. We also found continuum-like features for $q \in \{-\frac{\pi}{4}, \frac{7\pi}{4} \}$. 

In the non-interacting limit, $\lambda$ only splits the $j_{\textrm{eff}}=3/2$ and $j_{\textrm{eff}}=1/2$ bands, as shown in Fig.~\ref{fig6}(a), driving a metal to band-insulator transition. Only in the presence of finite $U$ the excitonic condensation happens: as shown in our DMRG results, $U/W \gtrapprox 1.0$ is required to obtain a noticeable staggering in the excitonic correlations. To further investigate the excitonic condensation at intermediate $U$, we calculated the single-particle spectral function $A_{jm}(q,\omega)$ (see Supplementary~\cite{splm}) at $U/W=2$, using both $\lambda=0.0$ and $\lambda/W=0.29$ corresponding to the IC-SDW and 
excitonic condensate region with $r_{coh} \approx 1.0$ 	
 (BCS limit), respectively. In Fig.~\ref{fig6}(b) we show $A_{jm}(q,\omega -\mu)$ for $\lambda=0$: at this point all $(j_{\textrm{eff}},m)$ states are degenerate. 
Comparing to the band structure in the non-interacting limit, a renormalization of the bands is clearly visible, having two minima structure and a local maxima at $q=\pi$, as a consequence of the Hubbard repulsion. In this phase, we expect that the nesting vector at the chemical potential $\mu$ decides the ordering vector of the 
incommensurate spin-density wave. We also show the single-particle density of states $\rho_{jm}(\omega-\mu)$, see Fig~\ref{fig6}(c), which indicates 
that the system has a finite density-of-states at $\mu$ suggesting this phase is metallic. 

Figures~\ref{fig6}(d) and (e) show 
$A_{j,m}(q, \omega - \mu)$ for $j_{\textrm{eff}}=3/2$ (all $m \in \{\pm 3/2, \pm 1/2 \}$ are degenerate) and $j_{\textrm{eff}}=1/2$ (both $m \in \{\pm 1/2 \}$ are degenerate), respectively. The gaps at $\mu$, at wavevectors $q\approx0$ and $q\approx\pi$ in the spectral functions of $j_{\textrm{eff}}=3/2$ and $j_{\textrm{eff}}=1/2$, respectively, are clearly present. These gaps appear due to the formation of bound states of electrons and holes, arising from the $q \approx 0$ and $q\approx \pi$ states of the $j_{\textrm{eff}}=3/2$ and $j_{\textrm{eff}}=1/2$ bands, respectively. This leads to the creation of excitons with net momentum $\approx \pi$ (indirect excitons). We also noticed a non-trivial suppression of the spectral function 
near $q=\pi$ for $j_{\textrm{eff}}=3/2$, below $\mu$, but the explanation for these small features requires further work. 
However, it is evident that the gap opens due to the formation of indirect excitons and eventually leads to a BCS-like state. In Fig.~\ref{fig6}(f), the $j$-resolved 
density-of-states is shown to illustrate the suppression near $\mu$ for both the $j_{\textrm{eff}}=3/2$ and $1/2$ bands. A small but finite density-of-states at $\mu$ is present because of the broadening $\eta$ used. To confirm the transition from metal (at $\lambda=0$) to excitonic insulator (at $\lambda/W=0.29$), we performed finite-size scaling of the charge gap $\Delta_{c}=E_{G}(N+1) + E_{G}(N-1) - 2E_{G}(N)$ for both points, as shown in Fig.~\ref{fig6}(g). At $\lambda=0.29$, using 
system sizes $L=8,12,16,24,32,$ and $42$, we found the charge gap is quite robust 0.55~eV. Sizes
$L=8,16,20,28,32,$ and $40$ were used for the scaling at the $\lambda=0$ point, which indicates that $\Delta_{c}$ scales to $\approx 0$~eV in the 
thermodynamic limit.

Figure ~\ref{fig7} ends this section by displaying the full $\lambda$ vs $U$ phase diagram for our one-dimensional systems. The CPU-costly 
DMRG calculations were performed for all small circles shown in the phase diagram using a system size $L=16$. After obtaining the ground state, then 
spin-spin correlations, orbital-orbital correlations, and exciton-exciton correlations were calculated to analyze the properties of each phase. The $(j_{\textrm{eff}},m)$-resolved 
local electronic densities were also studied to identify the relativistic band insulator phase. The dashed line inside the excitonic condensate region (green region) is only a guide to the eyes to show the BCS and BEC limits of the excitonic condensate.

\section{Hartree-Fock results in two-dimensions}

\begin{figure}[!h]
\hspace*{-0.52cm}
\vspace*{0cm}
\begin{overpic}[width=1.0\columnwidth , height=1.2\columnwidth]{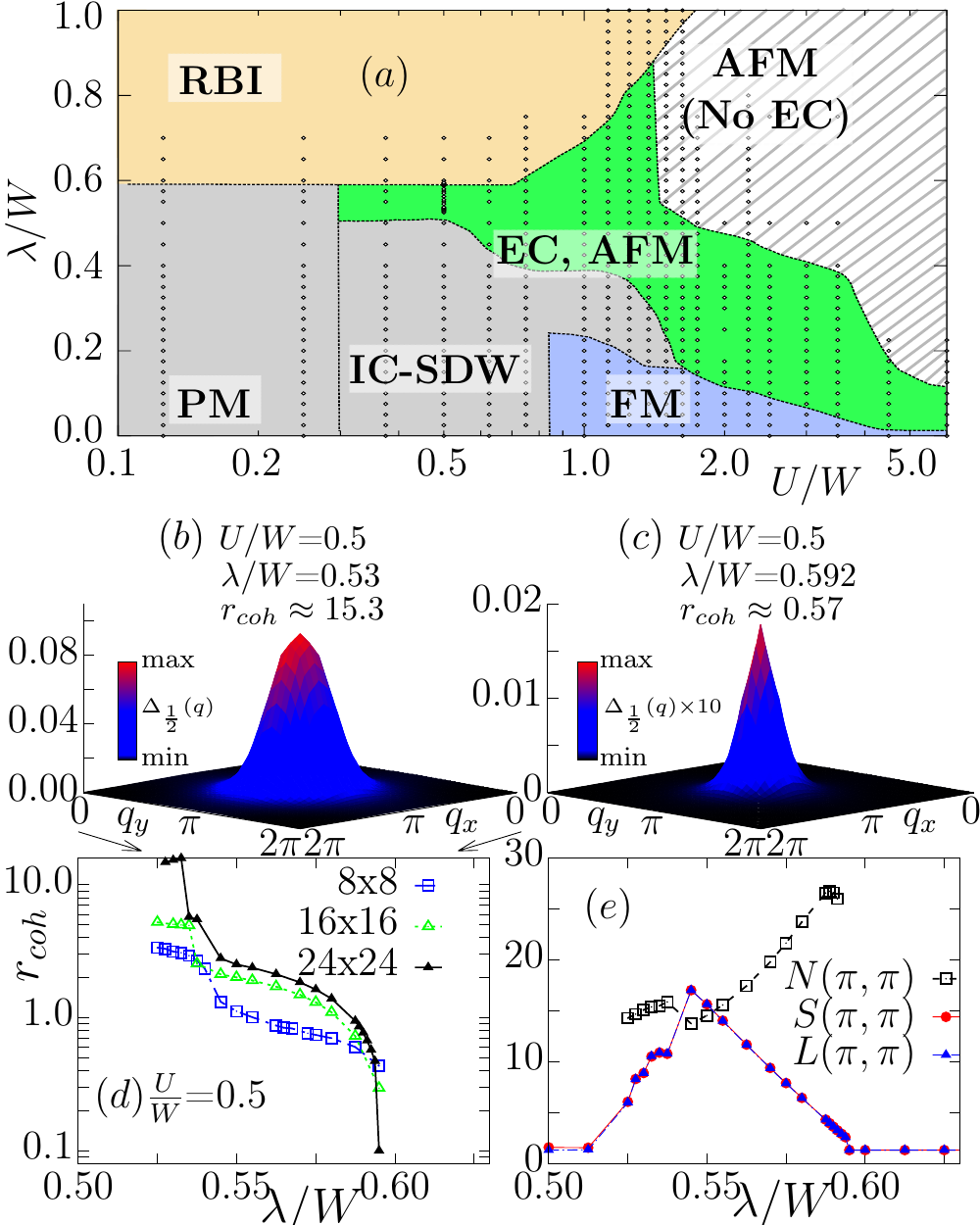}
\end{overpic}
\caption{Panel (a) shows the $\lambda$ vs $U$ phase diagram for the square lattice, calculated using the Hartree-Fock approximation. In panels (b) and (c), the  momentum distribution function of excitons $\Delta(\mathbf{q})$ is shown at $\lambda=0.53$ and $\lambda/W=0.592$, respectively, for fixed $U/W=0.5$ and a 
24$\times$24 system. The coherence length for various values of $\lambda$, at fixed $U/W=0.5$, for system sizes 8$\times$8, 16$\times$16, and 24$\times$24
are shown in panel (d). The $N(\pi,\pi)$, $S(\pi,\pi)$, and  $L(\pi,\pi)$ are presented in (e) for various values of $\lambda$, at fixed $U/W=0.5$ and using a 
24$\times$24 system size. }
\label{fig8}
\end{figure}

In this section, we will present and discuss the results obtained in two-dimensional lattices by performing mean-field calculations in real-space. All the four-fermionic terms in the Hubbard interaction Eq.(\ref{INT_term}) are treated under the Hartree-Fock approximation, where the single-particle density matrix expectation values $\langle c_{i\alpha\sigma}^{\dagger} c_{i\beta\sigma'}\rangle$ serve as order parameters at each site $i$ ($\alpha,\beta$ are orbitals and $\sigma,\sigma'$ are spin projection indexes). We reach self-consistency iteratively in the order parameters while tuning the chemical potential $\mu$ accordingly 
to attain the required electronic density. Given the many order parameters, converged results often require using 10-20 initial random initial configurations
and inspecting the lowest energy achieved after the iterative process, searching for patterns that are then uniformized and test for their energy. Often
also converged results at other couplings are used as seeds at new couplings, until a consistent phase diagram emerges. The modified Broyden's method was used to gain fast convergence~\cite{Johnson01}.

In Fig.~\ref{fig8}(a), we show the full $\lambda$ vs $U$ phase diagram for the two-dimensional lattice. To identify the various phases, we calculated the spin structure factor and orbital structure factor on 16$\times$16 clusters with periodic boundary conditions for all the points explicitly shown in the phase diagram. Our mean-field calculations in two dimensions capture almost all the phases found in our numerical exact one-dimensional results, the main difference being having shifted boundaries which are to be expected considering the different dimensionality and different many-body approximations. The only notable difference is that our mean-field calculations do not capture the non-magnetic phase in the strong coupling limit, because the lattice non-magnetic state 
can be written as a direct product of $J_{\textrm{eff}}=0$ at each site i.e. ... $|J_{\textrm{eff}=0}\rangle_{i}\otimes|J_{\textrm{eff}=0}\rangle_{i+1}\otimes|J_{\textrm{eff}=0}\rangle_{i+2}...$, where each $J_{\textrm{eff}}=0$ state in the $LS$ coupling limit is a sum of Slater determinants and hence cannot be reproduced by the Hartree-Fock approximation that relies on single determinants. However, the excitonic condensate phase, the focus of the present paper, is properly captured by the Hartree-Fock calculations and it is present even in a larger region of the phase diagram than in one dimension, hence giving us a good opportunity to discuss 
the presence of the BCS-BEC crossover in two-dimensional lattices.

To proceed with our discussion, we fix $U/W=0.5$ ($W=8t$) where the excitonic condensation is present in a narrow but finite range of spin-orbit coupling, while 
for smaller $\lambda$'s the IC-SDW phase is present. Similar to our one-dimensional DMRG calculations, in two-dimensional lattices we found that inside the excitonic condensate region (at a fixed weak or intermediate $U/W$ values), $r_{coh}$ decreases on increasing $\lambda$, as shown in Fig.~\ref{fig8}(d) depicting the BCS-BEC crossover. We have calculated $r_{coh}$ for system sizes 8$\times$8, 16$\times$16, and 24$\times$24. We found that in the BCS limit, $r_{coh}$ increases as the system size increases and $r_{coh}$ can reach values as high as $\approx 15.0a$ for the 24$\times$24 lattice. This clearly supports our claim for the presence of the BCS state above the IC-SDW region, as in our DMRG chain calculations (but in one-dimension perhaps due to size-effects or lack of resolution, the largest $r_{coh}$ obtained 
was only of order of one lattice spacing). On the other hand, we found that in the BEC limit, located below the relativistic band insulator in the phase diagram, 
$r_{coh}$ is $\mathcal{O}(0.1a)$, as shown in Fig.~\ref{fig8}(d). Figure~\ref{fig8}(e) displays $S(\pi,\pi)$ and $L(\pi, \pi)$, for $U/W=0.5$,  to show that only for a finite range of $\lambda$ the antiferromagnetic ordering develops. We also show the momentum distribution function of 
excitons $\Delta_{1/2}(\bold{q})$ at $\lambda/W=0.53$ and $\lambda/W=0.592$ in the BCS and BEC limits, respectively. We noticed that in the BEC limit $\Delta(\bold{q})$ is much sharper near $\bold{q}=(\pi,\pi)$ than in the BCS limit, as expected because in BEC a larger ratio of excitons is expected at the condensation momentum than other momenta. To further investigate the above claims we calculated the ratio of excitons at wavevector $\bold{q}=(\pi,\pi)$ and at other wavevectors using $N(\pi,\pi) = \Delta_{1/2}(\pi,\pi)/\langle \Delta_{1/2}(\bold{q}\ne (\pi,\pi)) \rangle$, where $\langle \Delta_{1/2}(\bold{q}\ne (\pi,\pi)) \rangle = \frac{1}{L^2 - 1}\sum_{\bold{q}\ne (\pi,\pi)} \Delta_{1/2}(\bold{q})$. It is evident from Fig.~\ref{fig8}(e) that $N(\pi,\pi)$ increases as we transition from the BCS to the BEC limits.

\begin{figure}[!t]
\hspace*{-0.52cm}
\vspace*{0cm}
\begin{overpic}[width=1.0\columnwidth , height=1.2\columnwidth]{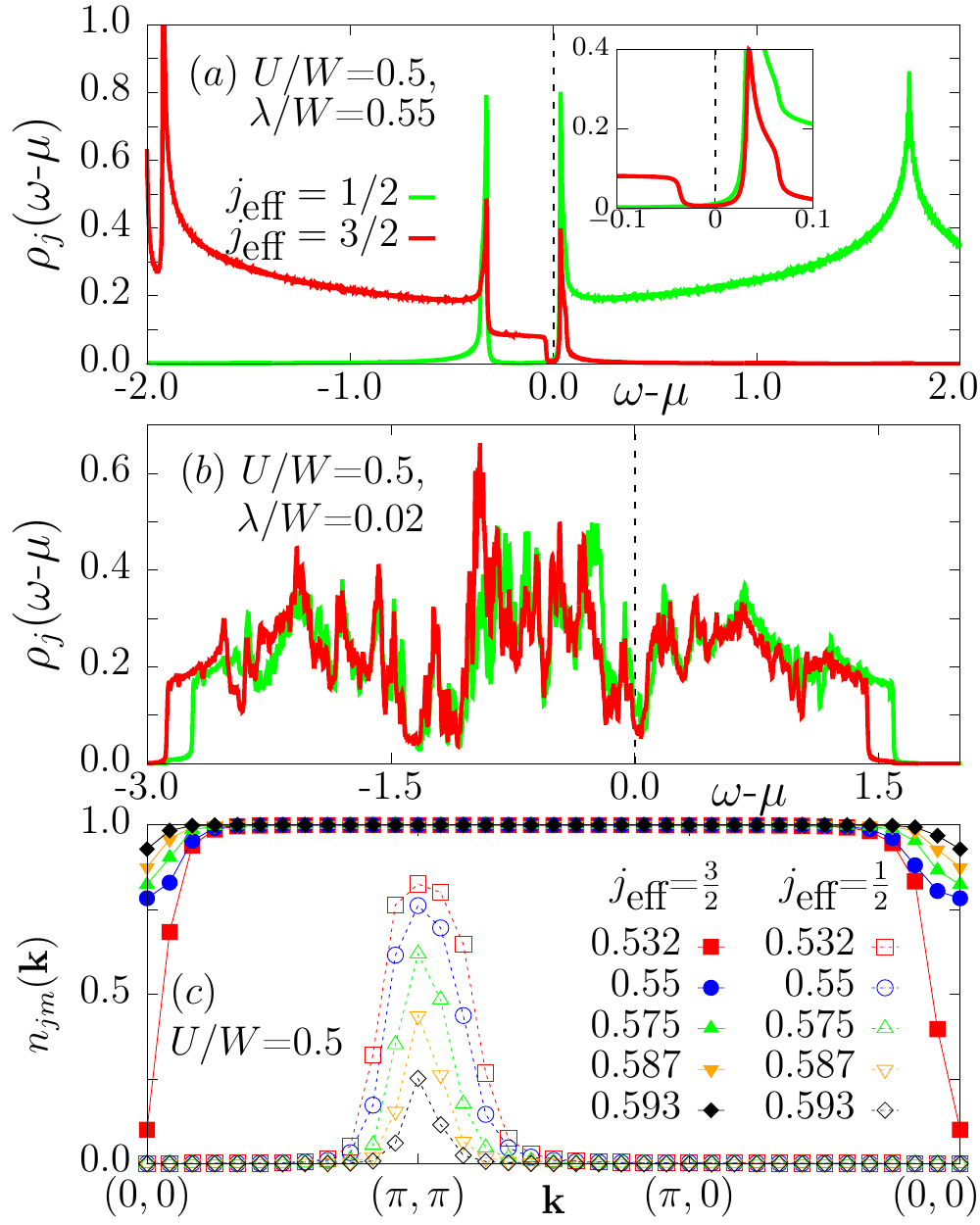}
\end{overpic}
	\caption{The density of states, $\rho_{j}(\omega -\mu)$, in the Excitonic Insulator and IC-SDW regions is shown in panels (a) and (b), respectively. Panel (c) contains the momentum distribution funtion for electrons $n_{jm}(\mathbf{k})$ for various values of $\lambda$. $U/W=0.5$ is fixed for all the results in the panels above. 
In panel (c), a system size 24$\times$24 is used and $m=1/2$ is fixed.}
\label{fig9}
\end{figure}

The single-particle density of states (DOS) provides further evidence
for the existence of the excitonic condensate. Figure ~\ref{fig9}(a) and (b) show the DOS $\rho_{j}(\omega - \mu)$, at fixed $U/W=0.5$. We chose $\lambda/W=0.02$ and $\lambda/W=0.55$ corresponding to the IC-SDW and excitonic condensate phases, respectively. As shown in Fig.~\ref{fig9}(b), at $\lambda/W=0.02$ a finite density-of-states at the chemical potential is present for both bands $j_{\textrm{eff}}=3/2$ and $j_{\textrm{eff}}=1/2$ indicating that the system is metallic in the IC-SDW phase. In the excitonic condensate phase, a small gap is present near the chemical potential in both the bands due to the condensation of excitons. The continous distribution in eigenenergies was attained by solving 24$\times$24 lattices with 48$\times$48 twisted boundary conditions~\cite{Salafranca01}. Two peaks arising from the small hole pocket of the $j_{\textrm{eff}}=1/2$ states and electron pocket ot the $j_{\textrm{eff}}=3/2$ states are also visible. The inset shows the magnified $\rho_{j}(\omega - \mu)$ near the chemical potential. To discuss the evolution of these hole and electron pockets, as the system crossovers from the BCS to BEC limit, we show the momentum distribution function of electrons $n_{jm}({\bf k})$ in Fig.~\ref{fig9}(c). The hole pocket is present at $\bold{k}=(0,0)$ in the  $j_{\textrm{eff}}=3/2$ band and the electron pocket is at $\bold{k}=(\pi,\pi)$ in the $j_{\textrm{eff}}=1/2$ band. The nesting vector between these pockets also explains the ordering momentum of the excitons. As $\lambda$ increases, we noticed a gradual decrease in both electron and hole pockets which indicates that the 
number of carriers decreases as the system crossovers from the BCS to BEC limits.

\section{Conclusions}
In this publication, we studied the degenerate $t_{2g}^{4}$ multiorbital Hubbard model in the presence of spin-orbit coupling, using one-dimensional chains and  numerically exact DMRG and also using two-dimensional clusters within the Hartree-Fock approximation. In both calculations, we provide evidence for a BCS-BEC crossover in the spin-orbit excitonic condensate, in the regime where the Hund coupling $J_{H}$ is fixed at a  robust value $J_H=U/4$. Within our accuracy, we established that in this model and at intermediate $U/W$, the system transits from an IC-SDW metallic phase to the BCS limit of an antiferromagnetic excitonic condensate, and on further increasing $\lambda$ the coherence length of electron-hole pairs decreases rapidly as the system crossovers to the BEC regime. This BEC regime ends as eventually the system transits to the relativistic band insulator on increasing further $\lambda$. Our work provides the  first indications of a BCS-BEC crossover in the excitonic 
magnetic state at intermediate $U/W$, a region of couplings that cannot be explored within approximations developed in the large $U/W$ regime.

We hope our study encourages further theoretical and experimental investigations on $t_{2g}^{4}$ coumpounds with robust spin-orbit coupling. Although our study is performed using degenerate orbitals, we believe that our findings could be generic and relevant for materials showing magnetic excitonic condensation 
at intermediate values of the Hubbard repulsion and spin-orbit coupling.

\section{Acknowledgments}
N.K., R.S., and E.D. were supported by the U.S.
Department of Energy (DOE), Office of Science, Basic Energy Sciences (BES), Materials Sciences and Engineering
Division. A.N. was supported by the Canada First Research Excellence Fund. G.A. was supported by the Scientific Discovery through Advanced Computing (SciDAC) program funded by the U.S. DOE, Office of Science, 
Advanced Scientific Computing Research and Basic Energy Sciences, Division of Materials Sciences and Engineering. Part of this work 
was conducted at the Center for Nanophase Materials Sciences, sponsored by the Scientific User Facilities Division (SUFD), BES, DOE, under contract with 
UT-Battelle.

\FloatBarrier


\end{document}